\documentclass[preprint,showpacs,preprintnumbers,amsmath,amssymb]{revtex4}

\usepackage{graphicx}
\usepackage{dcolumn}
\usepackage{bm}

\begin{document}


\title{Gravitomagnetism in Teleparallel Gravity}

\author{E. P. Spaniol}
\author{V. C. de Andrade}%
\email{andrade@fis.unb.br} \affiliation{Instituto de F\'{\i}sica,
Universidade de Bras\'{\i}lia
\\ C. P. 04385, 70.919-970 Bras\'{\i}lia DF, Brazil}

\date{\today}

\begin{abstract}
The assumption that matter charges and currents could generate
fields, which are called, by analogy with electromagnetism,
gravitoeletric and gravitomagnetic fields, dates from the origins of
General Relativity (GR). On the other hand, the Teleparallel
Equivalent of GR (TEGR), as a gauge theory, seems to be the ideal
scenario to define these fields, based on the gauge field strength
components. The purpose of the present work is to investigate the
nature of the gravitational electric and magnetic fields in the
context of the TEGR, where the tetrad formalism behind it seems to
be more appropriated to deal with phenomena related to observers.
 As our main results, we have obtained, for the first time,
the exact expressions for the gravito-electromagnetic fields for the
Schwarzschild solution that in the linear approximation become the
usual expected ones. To improve our understanding about these
fields, we  have also studied the geometry produced by a spherical
rotating shell in slow motion and weak field regime. Again, the
expressions obtained are in complete agreement with those of
electromagnetism.
\end{abstract}

\maketitle

\section{Introduction}

\indent

Gravitomagnetism has a history that is at least as long as that of
GR itself. In fact, it comes from the formal analogy between
Newton's law of gravitation and Coulomb's law of electromagnetism.
Both theories are governed by the same geometric law in a static
scheme and the interactions they describe propagate at finite speed.
Hence, the concept of a gravitoelectric description of the Newtonian
law is quite direct, and from it emerges the intuitive idea that
moving masses might generate the gravitational analogue of magnetic
fields.

The Maxwell-type gravitational theory was firstly explored by
Maxwell himself \cite{Maxwell} and  subsequently by some authors in
the second half of the nineteen century
\cite{Holzmuller,Tisserand,Heaviside}. Einstein also worked on this
parallel concomitantly to the birth of GR \cite{Einstein} and soon
after the publication of GR and its prediction of a gravitomagnetic
field some astrophysical applications start to be investigated
\cite{desitter,LT}.

After that, several authors have been studying the Einstein's
equations when considering a perturbation on flat spacetime
resulting in what they call linearized GR. In this context, the
remarkable similarity between gravitational field equations and the
Maxwell's becomes evident and we see the theoretical origin of a
gravitomagnetic field associated to masses currents
\cite{authors_linearized}. This subject is now well established and
discussed in several books of area \cite{Rindler,Sciama,wheeler}.

The gravitomagnetic 'dragging of inertial frames' by rotating matter
has played an important part in discussions about the meaning and
usefulness of Mach's principle, in astrophysical models of jets near
accreting, rotating black holes and in proposals for testing
alternative theories of gravity. All these aspects justify the
strong experimental efforts during the past 30 years to measure
gravitomagnetism. These attempts have been, however, hampered by a
intrinsic difficulty related to the fact that the gravitomagnetic
contribution is much smaller than the gravitoelectric one. In fact,
the Lense-Thirring precession of planetary orbits is too weak to be
measurable at present. On the other hand, since the early 1960's,
the measurement of the precession of a gyroscope has been the goal
of the Gravity Probe B experiment \cite{GP-B} and the evidence of
the gravitomagnetic field of the Earth (measuring nodal precession)
has been offered by Ciufolini et. al \cite{ciufolini} by studying
the motion of the laser-ranged satellites LAGEOS, LAGEOS II and
LAGEOS III. Measurements of the gravitomagnetic field around
superconductors seem to show its first signs \cite{super}.

Within the framework of GR, gravitomagnetism usually allows two
different theoretical approaches, which were summarized by Mashhoon
in \cite{mashhoon}. The first one is the context of linearized GR,
mentioned above, where it is obtained  essentially the analogous
equations to Maxwell's ones in the linear approximation, that is, by
performing an expansion in the metric tensor
$g_{\mu\nu}=\eta_{\mu\nu}+h_{\mu\nu}$, where $\eta_{\mu\nu}$ is the
Minkowski metric and $h_{\mu\nu}$ is the perturbation term. Then,
the gravitational potentials $\Phi$ and $\vec{A}$ are identified as
$\Phi=c^2\overline{h}_{0 0}/4$ and $A_i=-c^2\overline{h}_{0 i}/2$,
with $\overline{h}_{\mu \nu}={h}_{\mu \nu}-\frac{1}{2}\eta_{\mu
\nu}h$ and $h=\eta^{\mu \nu}h_{\mu \nu}$, and they in turn define
the physical fields. In the second approach, Mashhoon uses Fermi
coordinates to write the gravitoelectric and gravitomagnetic fields
as components of the curvature tensor. These coordinates, according
to Synge, are the correct relativistic generalization of the
Newtonian concept of reference frame \cite{synge}. Finally, we can
mention other formulation of GR known as Quasi-Maxellian, that is
based on a dynamics to Weyl tensor \cite{lich,ellis,maartens}.
Despite being a fully covariant formalism, it fails precisely
because it compares completely different objects \cite{goulart}.

In the present work, a different approach will be adopted to
reexamine gravitomagnetism. Due to the fundamental character of the
geometric structure underlying gauge theories, the concept of
charges and currents and, in particular, the concept of energy and
momentum are much more transparent when considered from a gauge
point of view \cite{duality}. Accordingly, we shall consider gravity
to be described as a gauge theory for the translation group
\cite{lorentz}, which gives rise to the so-called teleparallel
equivalent of GR. In this scenario we recover all the aspects
predicted by GR and moreover we have all the formal structure of a
gauge theory, which is naturally close to electromagnetism due to
its abelian character. Therefore, the concepts of gravitoelectric
and gravitomagnetic fields emerge, as we will see, in the same way
as in the electromagnetic theory, that is, as components of the
field strength of the gauge theory. Indeed, the concept of the
tetrad fields, that emerges from the gauge theory, and its relation
with observers in spacetime  \cite{Maluf}, results in the ideal
scenario to describe gravitoelectric and gravitomagnetic
observers-dependent fields.

Finally, we can say that one of our main results, obtained for the
first time, are the exact expressions for the
gravito-electromagnetic fields in a particular static distribution
of matter (Schwarzschild solution). This is a new approach, since
the calculations found in literature usually assume some
approximation hypothesis.

The paper is divided as follows: in section 2 the gravitational
Maxwell equations are introduced in their exact form. The next two
sections are devoted to applications of our definitions. Therefore,
in section 3 we study the exact and approximated Schwarzschild
solution and after that, in section 4, we apply our definition to
another case, that is, a spinning massive spherical shell in the
linearized approximation. Finally, in section 5 we draw the main
conclusions of the paper.

Notation: According to its gauge structure, to each point of
spacetime there is attached a Minkowski tangent spacetime (the fiber
of the correspondent tangent bundle), on which the translation
(gauge) group acts, and whose metric is assumed to be
$\eta_{ab}=(+1,-1,-1,-1)$. The spacetime indices will be denoted by
the Greek alphabet $(\mu,\nu,\sigma,...$ $= 0,1,2,3)$ and the
tangent space indices will be denoted by the first half of the Latin
alphabet $(a,b,c.. = 0,1,2,3)$.  The second half of the Latin
alphabet will be used to represent space tensor components, that is,
$(i,j,k...)$ assume the values 1,2 and 3. Indices in parentheses
will also be related to tangent space. We adopt the light velocity
as $c=1$.

\subsection{Teleparallel Gravity: a few concepts}

\indent

Let us present some of the more important expressions in
teleparallel gravity that will be used in the whole paper (for
detailed of the teleparallel fundamentals, see for example
\cite{lorentz,PRL}.

The field strength of the theory is defined in the usual form
\begin{equation}
F^{a}{}_{\mu \nu }=\partial _{\mu }A^{a}{}_{\nu }-
\partial_{\nu}A^{a}{}_{\mu } = h^{a}{}_{\rho }\;T^{\rho }{}_{\mu \nu } \;,  \label{fb}
\end{equation}
with $A^{a}{}_{\mu }$ the translational gauge potential and
$h^{a}{}_{\mu }=\partial_{\mu}x^a+A^{a}{}_{\mu }$ the tetrad field.
$T^{\rho }{}_{\mu \nu }$ is the torsion that represents alone the
gravitational field (in opposition to the curvature, that vanishes
in the Weitzeb\"ock space) and it also can be identified as the
field strength written in the tetrad base.

The dynamics of the gauge fields will be determined by the
lagrangian
\begin{equation}
\mathcal{L}_{G}=\frac{h}{16\pi G}\;S^{\rho \mu \nu }\;T_{\rho \mu
\nu }\;,
\end{equation}
with $h=\mathrm{det}(h^{a}{}_{\mu })$ and
\begin{equation}
S^{\rho \mu \nu }=-S^{\rho \nu \mu }\equiv {\frac{1}{2}}\left[
K^{\mu \nu \rho }-g^{\rho \nu }\;T^{\theta \mu }{}_{\theta }+g^{\rho
\mu }\;T^{\theta \nu }{}_{\theta }\right] \label{superpotencial}
\end{equation}
which is called superpotential, that will play an important role in
theory, as we will see.

The field equations resulting from this lagrangian are
\begin{equation}
\partial _{\sigma }(hS_{a}{}^{\sigma \rho }) - 4\pi G(hj_{a}{}^{\rho })=0  \label{EQcampo}
\end{equation}
with
\begin{equation}
j_{a}{}^{\rho }\equiv \frac{\partial{\mathcal
L}}{\partial{h^{a}{}_{\rho}}} =\frac{h_{a}{}^{\lambda}}{4\pi
G}(F^{c}{}_{\mu\lambda}S_{c}{}^{\mu\rho} - \frac{1}{4}
\delta_{\lambda}{}^{\rho}F^{c}{}_{\mu\nu}S_{c}{}^{\mu\nu}),
\label{ptemp2}
\end{equation}
$j_{a}{}^{\rho }$ stands for the gauge energy-momentum current of
the gravitational field.

\section{Gravitational Maxwell equations}

\indent

The TEGR, as mentioned in the last section, is an approach to
gravitation formulated as an abelian gauge theory, in the same sense
as electromagnetism, but associated with a different gauge group,
the translational one (in contrast with the U(1) group of
electromagnetic theory).

Our idea is to introduce a new version of gravitoelectric and
gravitomagnetic fields, by \text {straight} analogy with the
electric and magnetic fields of electromagnetism. These fields will
be proposed based on the gauge theory used, which means that, they
emerge in a completely different way that the
gravitoelectric/magnetic versions introduced in the GR scenario
\cite{mashhoon} and therefore the direct comparison can not be
performed in a simple manner.

After defining the fields we proceed in the opposite way taken in
electromagnetism and, from the covariant teleparallel form of the
field equations, we write them in a non covariant form similar to
Maxwell`s one that emphasize the phenomenology of what we call the
physical fields $\vec{E_{a}}$ and $\vec{B_{a}}$.

The immediate attempt to construct the gravitoelectric and
gravitomagnetic fields from this theory would consist in following
the usual gauge approach  and constructing the fields as components
of the field strength of the theory, which  means defining
$F_{a}{}^{0 i}=E_{a}{}^{i}$ and $F_{a}{}^{i j}=\epsilon^{i j k}B_{a
k}$. However, we investigate a more carefull assumption, that takes
into account the peculiar character of typical gauge theories for
gravitation: the possibility of contracting internal (algebra)
indices with external (spacetime) ones. Technically, this is
ascribed to the presence of a solder form, whose components
constitute the tetrad field. This property gives rise to deep
changes, contrasting with the usual internal (that is, non-soldered)
gauge theories. In teleparallelism, this effect was well observed,
for instance, in the construction of the lagrangian, which acquires
some additional terms in comparison with the usual gauge lagrangians
(see \cite{duality,lorentz}). The direct consequence of this
property, which finds echo in the field equations, is that the
quantity playing the role of the field strength, analogous to what
happens in Yang--Mills equations, will be, instead, the
generalization of the usual field strength, the superpotential
$S^a{}_{\mu \nu}$, given by (\ref{superpotencial}). It is natural,
therefore, to consider the gravitoelectric and gravitomagnetic
fields as components of this generalized field strength, that is,

\begin{equation}
S_{a}{}^{0 i}=E_{a}{}^{i},\label{definicao1}
\end{equation}
\begin{equation}
S_{a}{}^{i j}=\epsilon^{i j k}B_{a k}.\label{definicao2}
\end{equation}

Although these are the most appropriate definitions following the
gauge theories arguments, other approaches on teleparallel gravity
are present in literature \cite{APV,malufgrav,Blagojevic}.

\subsection{The first pair of field equations}

\indent

Let us consider the teleparallel version of the field equations for
gravitation outside the sources (in vacuum). They are given by
(\ref{EQcampo}):
\begin{eqnarray*}
\partial _{\sigma }(hS_{a}{}^{\sigma \rho })-4\pi G(hj_{a}{}^{\rho })=0.
\end{eqnarray*}

As these are the dynamical equations of the theory, we hope to
obtain the analogue of the first pair of Maxwell's equations of
electromagnetism. Hence, for $\rho=0$, we expect to find the
equivalent to Gauss's law and for $\rho=q$, we expect to obtain the
analogue of Amp\`ere's Law (with Maxwell's correction).

The field equation for $\rho=0$, when the model is applied, assumes
the general form:
\begin{equation}
\partial _{i}(hE_{a}{}^{i}) = 4\pi G(hj_{a}{}^{0}).
\label{eqj0}
\end{equation}

We thus see that the choice of the superpotential $S_{a}{}^{\sigma
\rho}$ as the generalized field strength of the theory leads to the
analogue of the divergent of $\vec{E}$ in the equation, up to a
multiplicative factor given by the determinant of the tetrad field.
In writing it in this form, we can directly interpret $hj_{a}{}^{0}$
as the source of the gravitoelectric field, in accordance with
Gauss's Law. This leads to the idea that 'gravitation generates
gravitation'.

Alternatively, we can replace our definitions (\ref{definicao1}) and
(\ref{definicao2}) in the gauge energy-momentum current
(\ref{ptemp2}), explicitly showing all the field contributions to
(\ref{eqj0}). This form makes evident the non-linearity of gravity,
that are manifested in the quadratic terms in the gravitoelectric
and gravitomagnetic components, as well as in the appearing crossing
terms. This complete form are shown in appendix A.

Let us consider now the spatial component of the field equation
$\rho=q$. We can write this equation in a compact form:
\begin{equation}
\epsilon^{q j k}\partial_{j}(hB_{a k})-\partial
_{0}(hE_{a}{}^{q})=4\pi G(hj_{a}{}^{q}). \label{campo2}
\end{equation}
In the same manner, $hj_{a}{}^{q}$ can be interpreted as the source
of the equation, in a very similar way to Amp\`ere's Law (with
corrections) of electromagnetism. The equation can also be written
in a expanded form, exhibited appendix A.

Hence, the first pair of field equations, in its exact form, is
evidently more complex than the corresponding electromagnetic pair.
This was in fact expected, due to the non-linearity of gravitation.
For some more interesting discussions about the these field
equations in the context of gravitational waves, see
\cite{nonlinear}.

\subsection{The second pair}

\indent

Analogously to electromagnetism, the second pair of gravitational
Maxwell equations is expected to emerge from the geometric context
of the gauge theory. Hence, we shall consider the first Bianchi
identity of teleparallel gravity, given by \cite{weitz}:
\begin{equation}
\partial _{\rho }F^{a}{}_{\mu \nu}+\partial _{\mu }F^{a}{}_{\nu
 \rho}+\partial _{\nu }F^{a}{}_{\rho \mu}=0. \label{Ibianchi}
\end{equation}
These equations ought to result in an equivalent of the second pair
of electromagnetic Maxwell equations.

We shall start from the relation between the field strength and the
superpotential, written in the form
\begin{eqnarray} \nonumber
F^{a}{}_{\gamma \delta} &=& 2h^{b}{}_{\gamma}g_{\rho
\delta}h^{a}{}_{\mu}S_{b}{}^{\mu \rho}- 2h^{b}{}_{\delta}g_{\nu
\gamma}h^{a}{}_{\mu}S_{b}{}^{\mu \nu} \\ &-& h^{a}{}_{\delta}g_{\nu
\gamma}h^{b}{}_{\theta}S_{b}{}^{\theta\nu}  +
h^{a}{}_{\gamma}g_{\rho \delta}h^{b}{}_{\theta }S_{b}{}^{\theta
\rho}
\end{eqnarray}
and introduce it directly into the teleparallel Bianchi identities
(\ref{Ibianchi}). Substituting the model definition, we get the
following equation:
\begin{eqnarray} \nonumber
&\partial_{\sigma}\left[{\cal O}^{b a}{}_{\gamma i
\delta}E_{b}{}^{i}+{\cal P}^{b a}{}_{\gamma i j \delta}\epsilon^{i j
k}B_{b k}\right] \\ &+ \nonumber
\partial_{\gamma}\left[{\cal Q}^{b a}{}_{\delta i \sigma}E_{b}{}^{i}+{\cal R}^{b
a}{}_{\delta i j \sigma}\epsilon^{i j k}B_{b k}\right] \\ & +
\partial_{\delta}\left[{\cal S}^{b a}{}_{\sigma i
\gamma}E_{b}{}^{i}+{\cal T}^{b a}{}_{\sigma i j \gamma}\epsilon^{i j
k}B_{b k}\right]=0.\label{EQcampoD2}
\end{eqnarray}

The first coefficients assume the following explicit form
\begin{eqnarray}\nonumber
{\cal O}^{b a}{}_{\gamma i \delta}&=&
2h^{b}{}_{\gamma}h^{a}{}_{0}g_{i
 \delta} - 2h^{b}{}_{\gamma}h^{a}{}_{i}g_{0 \delta}
- 2h^{b}{}_{\delta}h^{a}{}_{0}g_{i \gamma} +
2h^{b}{}_{\delta}h^{a}{}_{i}g_{0
 \gamma} \\ &+&
 h^{a}{}_{\delta}h^{b}{}_{0}g_{i
 \gamma}-h^{a}{}_{\delta}h^{b}{}_{i}g_{0 \gamma} -
h^{a}{}_{\gamma}h^{b}{}_{0}g_{i
 \delta}+h^{a}{}_{\gamma}h^{b}{}_{i}g_{0 \delta},
\end{eqnarray}
\begin{eqnarray}
{\cal P}^{b a}{}_{\gamma i j \delta} =
2h^{b}{}_{\gamma}h^{a}{}_{i}g_{j
 \delta} - 2h^{b}{}_{\delta}h^{a}{}_{i}g_{j \gamma} +
h^{a}{}_{\delta}h^{b}{}_{i}g_{j
 \gamma} - h^{a}{}_{\gamma}h^{b}{}_{i}g_{j \delta}.
\end{eqnarray}
From ${\cal O}^{b a}{}_{\gamma i \delta}$ and ${\cal P}^{b
a}{}_{\gamma i j
 \delta}$ we obtain ${\cal Q}^{b a}{}_{\delta i \sigma}$ and ${\cal R}^{b a}{}_{\delta i j
 \sigma}$, by switching the indices ($\gamma \rightarrow \delta$ and
 $\delta \rightarrow \sigma$) and ${\cal S}^{b a}{}_{\sigma i \gamma}$ and ${\cal T}^{b
 a}{}_{\sigma i j \gamma}$, by making ($\gamma \rightarrow \sigma$ and $\delta
 \rightarrow \gamma$).

We could, instead, try to make evident the analogy between these
equations and Maxwell's second pair of equations, performing the
substitutions $(\sigma \gamma \delta)\rightarrow (0 1 2),(0 1 3),(0
2 3)$, in order to obtain the analogue to Faraday's Law, or
performing the application $(\sigma \gamma \delta)\rightarrow(1 2
3)$ to get the gravitomagnetic divergence law. But, the equations,
in these exact forms, diverge from the electromagnetic ones
drastically.

Equations (\ref{EQcampo1D2}), (\ref{EQcampoD22}) and
(\ref{EQcampoD2}) acquire an aspect totally similar to the Maxwell
equations in the weak field limit, that is, when we compare
gravitation in a linear form and electromagnetism, a linear theory
in essence. It is explicitly shown in the next section.

Finally, it is interesting to note that the exact fields equations
in the usual GR gravitomagnetism, for an arbitrary curved spacetime,
as can be seen in Mashhoon's review \cite{mashhoon} also exhibit an
intricate form, but in that context the gravitoelectric and
gravitomagnetic fields are related to the $R_{0 i 0 j}$ and $R_{0 i
j k}$ Riemann tensor components. Only in the limit case (in lowest
order in $\left| \mathbf{X}\right| /\mathcal{R}$, with $\mathbf{X}$
the spacial components of the fermi coordinates and $\mathcal{R}$
the radius of curvature of spacetime), a kind of Maxwell's equations
are recovered.  We must emphasize that the conceptual definitions of
what would be the gravitoelectric and gravitomagnetic fields in
these two approaches are completely different, being the
teleparallel one much more consistent with electromagnetic gauge
theory definitions.

\section{Schwarzschild solution}

\indent

We have made, until now, a theoretical analogy between gravitation
and electromagnetism, which was based on the identification of the
superpotential components with the gravitoelectric and
gravitomagnetic fields. This parallel was drawn in the gauge
theories domain. The next step is to investigate whether these
fields are associated with physical phenomena similar to those
observed and expected in the context of electromagnetism. In short,
static electric charges generate electric fields and moving charges
are associated with magnetic fields. Similarly, it is desirable to
establish an association between the fields suggested in the model
and static and moving matter.

\subsection{Geometry}

\indent

Let us consider the geometry produced by the gravitational field of
a spherical symmetric and static distribution of matter, say, a body
represented by a mass m and situated at the origin of a coordinate
system. This is the Schwarzschild solution of Einstein's equations,
the most popular geometry in gravitation, since it allows us to
treat bodies like the sun and other celestial ones with excellent
approximation. It is given by the line element:
\begin{eqnarray}
ds^2&=&\left( 1-\frac{2mG}{r}\right)dt^2-\left(
1-\frac{2mG}{r}\right)^{-1}dr^{2} - r^{2}\left(d\theta^2+sin^2\theta
d\phi^2\right).
\end{eqnarray}

The choice of the observer, that corresponds to a specific tetrad
field associated with the Schwarzschild metric, is crucial to our
conclusions about the fields $\vec{E^{a}}$ and $\vec{B^{a}}$ that
emerge in theory, as in electromagnetic case. Hence, we adopt a set
of tetrads that is adapted to a stationary observer localized at
infinity. As such, the tetrad needs to satisfy the time gauge
condition and exhibit symmetry in the spatial sector \cite{Maluf}.

This set of tetrad fields can be represented by \cite{Zhang}.
\begin{equation}
h^{a}{}_{ \nu}=
\left(%
   \begin{array}{cccc}
          \gamma_{00} & 0& 0 & 0 \\
          0 & \gamma_{11} sin\theta cos\phi & rcos\theta cos\phi & -rsin\theta sin\phi\\
          0 & \gamma_{11} sin\theta sin\phi & rcos\theta sin\phi & rsin\theta cos\phi \\
          0& \gamma_{11}cos\theta & -rsin\theta & 0\\
   \end{array}%
\right).\label{tetradasch}
\end{equation}
In this notation, $\gamma_{00}=\sqrt{g_{00}}$ and
$\gamma_{11}=\sqrt{-g_{11}}$. From the usual expression for torsion
written in terms of the tetrad
\begin{equation}
T^{\sigma }{}_{\mu \nu }=h_{a}{}^{\sigma }\partial _{\mu }h^{a
}{}_{\nu } - h_{a}{}^{\sigma }\partial _{\nu }h^{a }{}_{\mu
},\label{Thh}
\end{equation}
we calculate the components of $T^{\sigma }{}_{\mu \nu }$, the
non-zero of which are given by:
\begin{equation}
T^{0}{}_{0 1 }= - \frac{GM}{r^{2}} g_{00}^{-1},
\end{equation}
\begin{equation}
T^{2}{}_{1 2 }= \frac{1}{r} (1-\gamma_{11}),
\end{equation}
\begin{equation}
T^{3}{}_{1 3 }= \frac{1}{r}(1-\gamma_{11}).
\end{equation}

Considering the definition (\ref{definicao1}), we obtain a direct
relation between the electric field and torsion:
\begin{equation}
E_{b}{}^{i}=\frac{1}{2}h_{b}{}^{0}T^{j i}{}_{j}.
\end{equation}

For $b\neq0$ it is trivial to verify that the components vanish,
that is
\begin{equation}
E_{(k)}{}^{i}=0
\end{equation}
and for $b=0$ we get
\begin{equation}
E_{(0)}{}^{i} = \frac{1}{2}h_{(0)}{}^{0}(T^{1 i}{}_{1}+T^{2
i}{}_{2}+T^{3 i}{}_{3}).
\end{equation}
Substituting explicitly the tetrad and torsion components, we obtain
\begin{eqnarray}
E_{(0)}{}^{r}&=&-\frac{1}{r}(\gamma_{00}-1), \\
E_{(0)}{}^{\theta}&=&0, \\
E_{(0)}{}^{\phi}&=&0.
\end{eqnarray}
This means that only the radial component of the vector $b = 0$ is
different from zero, something which is consistent with the
spherically symmetric distribution of mass. Considering
(\ref{definicao2}) we get
\begin{eqnarray}
S_{b}{}^{1 2}&=&-\frac{1}{2}h_{b}{}^{2}g^{1 1}[T^{0}{}_{1 0}+T^{3}{}_{1 3}], \\
S_{b}{}^{1 3}&=&-\frac{1}{2}h_{b}{}^{3}g^{1 1}[T^{0}{}_{1 0}+T^{2}{}_{1 2}], \\
S_{b}{}^{2 3}&=&0.
\end{eqnarray}
And finally, we obtain the gravitomagnetic components
\begin{eqnarray}\nonumber
B_{(0) \phi}&=&0,
\\ \nonumber
B_{(1) \phi}&=&\frac{cos\theta
cos\phi}{2r^2}(1-\gamma_{11}{}^{-1}-\frac{GM}{r}),
\\ \nonumber
B_{(2) \phi}&=&\frac{cos\theta
sin\phi}{2r^2}(1-\gamma_{11}{}^{-1}-\frac{GM}{r}),
\\ \nonumber
B_{(3)
\phi}&=&-\frac{sin\theta}{2r^2}(1-\gamma_{11}{}^{-1}-\frac{GM}{r}),
\\ \nonumber
B_{(0) \theta}&=&0,
\\ \nonumber
B_{(1) \theta}&=&\frac{sin\phi}{2r^{2}
sin\theta}(1-\gamma_{11}{}^{-1}-\frac{GM}{r}),
\\ \nonumber
B_{(2) \theta}&=&-\frac{cos\phi}{2r^{2}
sin\theta}(1-\gamma_{11}{}^{-1}-\frac{GM}{r}),
\\ \nonumber
B_{(3) \theta}&=& 0.
\end{eqnarray}

As can be easily seen, the scalar resulting from the contractions of
all internal and external indices that can be associated to a 'kind'
of modulus of a gravitomagnetic vector does not present angular
dependence, which is in agreement to the spherical symmetry of the
Schwarzschild solution, {that is, $B^2\equiv g_{ij}B_{a}{}^{i}
B^{a}{}^{j}=\frac{1}{gr^{2}}(1-\gamma_{11}{}^{-1}-\frac{GM}{r})^2$.

The appearance of non null components of gravitomagnetic components,
as shown above, can be associated with the gravitational
energy-momentum current $j_{a}{}^{\rho }$, that represents the non
linear effects of gravitation. In this way, the emergence of
gravitomagnetism even in static configurations is theoretically
consistent and this fact introduces new possibilities of observation
in strong field regimes. Finally, we can say that a closer analogy
between gravitation and electromagnetism, in the phenomenological
point of view, must be investigated when both interactions are
treated as linear, that is, when we perform the weak field
hypothesis in the gravitational interaction.

\subsection{A test of consistency: approximate Schwarzschild
solution}

\indent

Considering now this approximation hypothesis, the spacetime metric
will be decomposed into a trivial flat part, $\eta_{\mu \nu}$, plus
a perturbation generated by the presence of matter, $a_{\mu \nu}$,
that is
\begin{equation}
g_{\mu \nu}=\eta_{\mu \nu}+a_{\mu \nu}.
\end{equation}
In this specific case, it is convenient, as we will see, to work in
Cartesian coordinates, which will simplify substantially our
calculations. The correspondent tetrad will be also divided into a
trivial (diagonal) part, $H^{a}{}_{\mu}$, plus a contribution due to
the presence of matter, $U^{a}{}_{\mu}$, that is
\begin{equation}
h^{a}{}_{\mu}=H^{a}{}_{\mu}+U^{a}{}_{\mu}.\label{hHU}
\end{equation}

The weak field limit means, essentially, that we must discard the
terms of second order $(\frac{m}{r})^2=O(\epsilon^2)$, that is,
\begin{equation}
\left(\frac{m}{r} \right)^2 << 1
\end{equation}
and then we can work with the Taylor expanded version of the studied
expressions.

Hence, we get the tetrad fields (\ref{tetradasch}) in cartesian
coordinates in first order:
\begin{equation}
h^{a}{}_{\mu}=
\left(%
   \begin{array}{cccc}
          1-\frac{mG}{\xi^{\frac{1}{3}}} & 0 & 0 & 0  \\
          0 & 1+\frac{mGx^2}{\xi} & \frac{mGxy}{\xi} & \frac{mGxz}{\xi} \\
          0 & \frac{mGxy}{\xi}&  1+\frac{mGy^2}{\xi}& \ \frac{mGyz}{\xi}\\
          0 &  \frac{mGxz}{\xi} &  \frac{mGyz}{\xi} & 1+\frac{mGz^2}{\xi} \\
   \end{array}%
\right)
\end{equation}
with $\xi=(x^2+y^2+z^2)^\frac{3}{2}$. Notice that the tetrad above
reduces to the addition of a diagonal part (that corresponds to a
Minkowski space only in this coordinate system) and a non trivial
part that is due to the presence of matter.

We can therefore calculate the torsion written in terms of the
tetrad through (\ref{Thh}) and we get the following non null
components:
\begin{equation}
T^{0}{}_{0 1}=T^{2}{}_{1 2}=T^{3}{}_{1
3}=-\frac{mGx}{(x^2+y^2+z^2)^\frac{3}{2}},
\end{equation}
\begin{equation}
T^{0}{}_{0 2}=T^{1}{}_{2 1}=T^{3}{}_{2
3}=-\frac{mGy}{(x^2+y^2+z^2)^\frac{3}{2}},
\end{equation}
\begin{equation}
T^{0}{}_{0 3}=T^{2}{}_{3 2}=T^{1}{}_{3
1}=-\frac{mGz}{(x^2+y^2+z^2)^\frac{3}{2}}.
\end{equation}

Let us consider again our gravito-eletromagnetic definitions applied
to a region in space that satisfies the weak field approximation to
the Schwarzschild geometry and also expression
(\ref{superpotencial}). For the gravitoelectric components defined
in (\ref{definicao1}), we find the following non-null components:
\begin{equation}
E_{(0)}{}^{x_k}=\frac{mGx_k}{(x^2+y^2+z^2)^\frac{3}{2}},
\end{equation}
with $x_k$ the usual spacial cartesian coordinates. This result
seems to be really interesting: the components $a=0$ of
$\vec{E^{a}}$ play a role quite analogous to the coulombian electric
field of electromagnetism. The other components $a=i$ do not
contribute to the gravitoelectric field.

Now, looking for the gravitational magnetic components, we see that
in this order of approximation, they are not present for a static
distribution of matter, that is,
\begin{equation}
B_{(0)k}=B_{(i)j}=0
\end{equation}
supporting the straight analogy between the theories under this
condition. We can also observe that in this order of approximation
there is no gravitational current contribution, as discussed below.
This fact reinforces the argument given earlier that $j_{a}{}^{\rho
}$ is responsible for the appearance of the gravitomagnetic
components in the exact Schwarzschild solution. Furthermore, we can
say that these expressions are in total agreement to that obtained
in the linearized GR (for example, see \cite{wheeler}), but its
origins are conceptually quite different.

\subsection{Field equations in Schwarzschild geometry}

\indent

We have concluded until now that our model generates components for
$\vec{E^{a}}$ and $\vec{B^{a}}$ that are compatible with the
electromagnetic phenomenology in the case of the most simple example
of static distribution of matter, that is, the Schwarzschild
geometry, considering weak gravitation. The next step is to examine
what happens with the equations that describe these fields.

\subsubsection{The first pair of field equations}

\indent

The so called dynamical equations of the theory, which correspond to
the gravitational analogue of the first pair of Maxwell´s equations,
resulted in the exact form in the expressions (\ref{EQcampo1D2}) and
(\ref{EQcampoD22}), that are evidently more complex then the
respective equations of electromagnetism. Nevertheless, considering
the suggested approximation of weak field limit, we will see, as
follows, that the laws become extremely simple, reinforcing the
analogy.

Taking the weak field limit, it is trivial to show that all
contributions that come from $j_{a}{}^{\rho}$ in (\ref{EQcampo1D2})
and (\ref{EQcampoD22}) are of  $O(\epsilon^2)$ order, just remaining
the derivative term
\begin{equation}
\partial_{\sigma}(hS_{a}{}^{\rho \sigma })=0
\end{equation}
with $h=1$. For $\rho=0$ we get

\begin{equation}
\partial_{\sigma}(S_{a}{}^{0 i})=0
\end{equation}
which corresponds to

\begin{equation}
\vec{\nabla}\cdot{\vec{E_{a}}}=0
\end{equation}
and for $\rho=i$, we find

\begin{equation}
\partial_{0}(S_{a}{}^{0 j})+\partial_{i}(S_{a}{}^{i j})=0,
\end{equation}
that assumes the simple form
\begin{equation}
\vec{\nabla }\times {\vec{B_{a}}}=\frac{\partial{\vec{E_{a}}}
}{\partial t}.
\end{equation}
and which is, in Schwarzschild solution, identically satisfied.

\subsubsection{The second pair of field equations}

\indent

The teleparallel Bianchi identities, that give us the second
(geometrical) pair of field equations, when calculated in the exact
form, have also resulted in an intricate relation, with several
coupling terms,  unexpected in electromagnetism. But in the same
way, the weak field approximation can disappear with these spurious
contributions, resulting in the following equations. Let us
consider, for example, $\sigma=0$, $\gamma=1$ and $\delta=2$ in the
equation (\ref{EQcampoD2}). For $a=0$ we get
\begin{equation}
\partial_{x}E_{(0)}{}^{y}-\partial_{y}E_{(0)}{}^{x}=0.
\end{equation}
Taking into account the other possibilities, that is, ($\sigma=0$,
$\gamma=1$, $\delta=3$) and ($\sigma=0$, $\gamma=2$, $\delta=3$) we
obtain an analogous of Faraday's law with $a=0$
\begin{equation}
\overrightarrow{\nabla }\times \vec{E_{(0)}}= - \frac{\partial
\vec{B_{(0)}}}{\partial t}.
\end{equation}
For $a=1,2,3$ it is trivially satisfied. On the other hand, equation
(\ref{EQcampoD2}) with $\sigma=1$, $\gamma=2$ and $\delta=3$, would
correspond to
\begin{equation}
\overrightarrow{\nabla }\cdot\vec{B_{a}}=0,
\end{equation}
nevertheless, in this order $\vec{B_{a}}=0$, the equation becomes
identically null.

Finally, we would like to emphasize that these equations, also
reproduced in the linearized GR context, were obtained here from the
first principles, in a closer and legitimate analogy to
electromagnetism.

\section{The spinning massive spherical shell}

\indent

In the last section, we have presented a direct test for our model,
in which it was considered a static geometry (Schwarzschild
solution) both for exact and approximate solutions.

Our purpose is now to analyze the behavior of the model when applied
to another configuration of spacetime, where gravitomagnetic
components are expected even in the linearized geometry. Thus, we
shall consider the spacetime geometry associated with a spherical
mass shell in slow rotation and, in the same fashion performed
above, the terms of order superior to $\epsilon$ shall be discarded,
keeping in mind that $\epsilon= \frac{m}{r}$ . The greatest interest
in choosing this metric is that it resembles the region outside the
Kerr spacetime, with the advantage of having no singularities. It
displays therefore regular rotational effects and it can be easily
treated mathematically.

\subsection{The geometry}

\indent

The metric tensor that represents the spherical mass shell in
rotation was firstly introduced by Cohen in \cite{Cohen}, and reads,
in spherical coordinates $(r, \theta,\phi)$:
\begin{equation}
ds^2=-V^2dt^2+\psi^4\lbrack dr^2+r^2d\theta^2+
r^2\sin^2\theta(d\phi-\Omega dt)^2\rbrack\;. \label{701}
\end{equation}
It is important to remark that this is a solution of Einstein´s
equation to first order in $\Omega$, the angular velocity. In the
following terms, $r_0$ is the radius of the shell and $\alpha =
\frac{m}{2}$, with $m$ its mass.

Inside the shell, that is, for the region given by $r<r_0$, we have:
\begin{eqnarray}
V={{ r_0-\alpha}\over{r_0 + \alpha}},
\end{eqnarray}
\begin{eqnarray}
\psi&=& \psi_0 =1+ {\alpha \over r_0},
\end{eqnarray}
\begin{eqnarray}
\Omega&=&\Omega_0.
\end{eqnarray}
For $r>r_0$:
\begin{eqnarray}
V={{ r-\alpha}\over{r + \alpha}},
\end{eqnarray}
\begin{eqnarray}
\psi= 1+ {\alpha \over r},
\end{eqnarray}
\begin{eqnarray}
\Omega=\left(\frac{r_0\psi_0^2}{r\psi^2}\right)^3\Omega_0.
\end{eqnarray}

The constant $\Omega_0$ stands for the dragging angular velocity of
locally inertial observers inside the shell.

Again, the choice of observers is fundamental to our conclusions,
and so we shall adopt a tetrad field adapted to static observers at
spacelike infinity. Therefore, the following condition must be
satisfied \cite{Maluf1}:
\begin{equation}
h_{(0)}\,^\mu(t,r,\theta,\phi)= \left(\frac{1}{A},0,0,0 \right)\,.
\end{equation}
A tetrad field that meets this condition is, for example,
\begin{equation}
h^{a}{}_{\mu}=
\left(%
   \begin{array}{cccc}
          A& 0 & 0 & -C  \\
          0 & \psi^2 sin\theta\,\sin\phi & r\psi^2\cos\theta\,\cos\phi & -B sin\theta\, sin\phi \\
          0 & \psi^2 sin\theta\, sin\phi & r\psi^2 \cos\theta\, sin\phi & B sin\theta\,\cos\phi \\
          0 & \psi^2 \cos\theta  & -r\psi^2 sin\theta & 0  \\
   \end{array}\label{702}
\right)
\end{equation}
with
\begin{eqnarray}
A&=&(V^2-r^2 \Omega^2 \psi^4 sin^2\theta)^{1/2}\,, \nonumber \\
C&=& -{1\over A}\, \Omega r^2 \psi^4 sin^2\theta \,, \nonumber \\
B&=&{V\over A}\, r\psi^2\,.\label{703}
\end{eqnarray}

As the application of the weak field limit results in the simple
decomposition of a diagonal tetrad (corresponding to the flat
metric) plus a non-trivial term (due to the presence of mass and
rotation) only in Cartesian coordinates, we write (\ref{702}) in
this coordinate system as
\begin{equation}
h^{a}{}_{\mu}=H^{a}{}_{\mu}+U^{a}{}_{\mu}.\label{705}
\end{equation}

In order to obtain the expression above, it is necessary to make
some assumptions. The first one is the same used in the last
section, that means, to discard the terms of order
$\left(\frac{\alpha}{r}\right)^2=\epsilon^2$. We shall also take the
limit
\begin{equation}
\left(\frac{\alpha}{r_0}\right)^2<< 1,
\end{equation}
and therefore we can perform a Taylor expansion to obtain
expressions that have these terms. Moreover, we shall consider the
weak rotation regime, that is
\begin{equation}
r^2 \Omega^2 << 1.\label{rotacaolenta}
\end{equation}
Considering (\ref{rotacaolenta}), the definitions (\ref{703}) become
\begin{eqnarray}
A& \approx & V\,, \nonumber \\
C& \approx & -{1\over V}\, \Omega r^2 \psi^4 sin^2\theta \,, \nonumber \\
B& \approx & r\psi^2\,.\label{706}
\end{eqnarray}

After making these considerations, we can write the tetrad field in
Cartesian coordinates and finally get
\begin{equation}
h^{a}{}_{\mu}=
\left(%
   \begin{array}{cccc}
          1-\frac{2\alpha}{\chi} & -\frac{r_0^2(r_0 + 6\alpha) \Omega_0 y}{\chi^{3}} & \frac{r_0^2(r_0 + 6\alpha) \Omega_0 x}{\chi^{3}} & 0 \\
          0 & 1+\frac{2\alpha}{\chi} & 0 & 0 \\
          0 & 0&  1+\frac{2\alpha}{\chi} & 0 \\
          0 & 0 &  0 & 1+\frac{2\alpha}{\chi} \\
   \end{array}%
\right), \label{716}
\end{equation}
with
\begin{equation}
\chi=\sqrt{x^2+y^2+z^2}
\end{equation}
which corresponds to a composition of a flat tetrad and a
perturbation, as we wanted.

Thus we can calculate the torsion in first order substituting
(\ref{705}) in (\ref{Thh}). These expressions are in appendix B.

\subsection{Gravitomagnetic field of a rotating massive spherical shell}

\indent

Having all necessary elements, we can evaluate the gravitomagnetic
components of a gravitational field produced by a massive spherical
shell in slow rotation.

According to the definition (\ref{definicao2}), the $x$ component
 of the gravitomagnetic field becomes
\begin{eqnarray} \nonumber
B_{(0) x} &=&  \frac{1}{4} h_{(0)}{}^{0} g^{22}g^{33}\left(T_{023}+
T_{203}\right) + \frac{1}{2}[h_{(0)}{}^{0}g^{20}g^{33}T_{003}\\ &-&
h_{(0)}{}^{2}g^{00}g^{33}T_{003} - h_{(0)}{}^{2}g^{11}g^{33}T_{113}]
,
\end{eqnarray}
the $y$ component is
\begin{eqnarray} \nonumber
B_{(0) y} &=& - \frac{1}{4} h_{(0)}{}^{0} g^{11}g^{33}\left(T_{013}+
T_{103}\right)  - \frac{1}{2}[h_{(0)}{}^{0}g^{10}g^{33}T_{003}\\ &+&
h_{(0)}{}^{1}g^{00}g^{33}T_{003} +
h_{(0)}{}^{1}g^{22}g^{33}T_{223}],
\end{eqnarray}
and finally, the $z$ component is
\begin{eqnarray} \nonumber
B_{(0) z} &=& \frac{1}{2}[h_{(0)}{}^{2}g^{11}g^{00}T_{001}  -
h_{(0)}{}^{0}g^{11}g^{20}T_{001} +
h_{(0)}{}^{0}g^{10}g^{22}T_{002}\\ \nonumber &-&
h_{(0)}{}^{2}g^{11}g^{33}T_{313} - h_{(0)}{}^{1}g^{22}g^{00}T_{002}
+ h_{(0)}{}^{1}g^{22}g^{33}T_{323}]\\ &+& \frac{1}{4} h_{(0)}{}^{0}
g^{11}g^{22} \left[ - T_{201}+ T_{012} - T_{102}\right].
\end{eqnarray}

All other superpotential combinations are seen to be zero, taking
into account our hypothesis.

After substituting the tetrad elements, the metric tensor and
torsion components in the equations above, we obtain the following
gravitomagnetic $x$, $y$ and $z$ components, respectively,
\begin{eqnarray}
B_{(0) x} = \frac{r_0^2 \Omega_0(r_0 +
6\alpha)xz}{2\chi^5}\left[\frac{3}{2} - \frac{11\alpha}{\chi}
\right] \label{B1},
\end{eqnarray}
\begin{eqnarray}
B_{(0) y} = \frac{r_0^2 \Omega_0(r_0 +
6\alpha)yz}{2\chi^5}\left[\frac{3}{2} - \frac{11\alpha}{\chi}
\right]\label{B2}
\end{eqnarray}
and
\begin{eqnarray}
B_{(0) z} &=& \frac{r_0^2 \Omega_0(r_0 + 6\alpha)}{2\chi^3}\left[1 -
\frac{8\alpha}{\chi} \right] - \frac{r_0^2 \Omega_0(r_0 +
6\alpha)}{2\chi^5}(x^2 + y^2)\left[\frac{3}{2} -
\frac{11\alpha}{\chi} \right].\label{B3}
\end{eqnarray}
It is interesting to perform a comparison between our results and
the analogous problem in electromagnetism. Thus, let us consider a
spherical shell of radius $R$, bearing a superficial charge density
$\sigma$ and which is placed to move with an angular velocity
$\Omega$ \cite{Griffiths}. After some direct calculations, we obtain
the magnetic field of this configuration in Cartesian coordinates to
be
\begin{equation}
\vec{B} = \frac{\mu_0 R^4 \Omega \sigma}{3\chi^5} \left[3( xz\hat{x}
+ yz\hat{y}) - (x^2 + y^2 - 2z^2)\hat{z} \right]\label{704}
\end{equation}
where $\mu_0$ is the permeability constant.

Comparing the above expressions (\ref{B1}-\ref{B3}) with
(\ref{704}), we note a strong similarity between them. For example,
the $x$ and $y$ components of $B^{a}$ (a=0) are proportional to the
product $xz$ e $yz$, respectively, in the same way as the magnetic
ones, while the $z$ component has a term proportional to $x^2$ and
$y^2$, like in electromagnetism. In fact, we can easily see that the
expressions (\ref{B1}-\ref{B3}) correspond in zero order to a
magnetic dipole field. Finally, if we stop the rotation, by making
$\Omega_0$ equal to zero, we see that the above components will also
disappear, that is, if the shell loses its rotation motion,
gravitomagnetism disappears. Again we can say that our results are
consistent with similar situations studied in literature in the GR
version of gravitomagnetism \cite{wheeler} despite using conceptual
different definitions of what would be gravitomagnetic field, being
the teleparallel ones much more appropriate to draw comparisons with
electromagnetism.

\section{Conclusions}

\indent

Teleparallel gravity, though equivalent to GR, has its fundamentals
in an abelian gauge theory, in the same way as electromagnetism.
This motivates an attempt to put the dynamical equations in a
similar form and to make a closer interpretation of its
phenomenology. As the crucial differences between these two
fundamental interactions of nature are present, the parallel was
carried out as far as possible, as we summarize below.

Those which would be the gravitational Maxwell equations in their
exact form proved to be definitely more complex than the
electromagnetic ones. This was in fact an expected result, which has
its origin in the strong non linearity of gravity. The derivatives
of the gravitoelectric and gravitomagnetic fields appearing in the
gravitational equations are in general similar to the
electromagnetic ones, but we see the emergence of coupling fields
terms playing the role of source of gravitation, which characterizes
what we have identified as  ´gravitation as source of gravitation´.
The exact form of the field equations in the standard GR
gravitomagnetism, despite conceptually different, also presents the
same complexity.

In an attempt to understand whether the model proposed was
associated with physical phenomena analogous to the electromagnetic
ones, we first applied our definitions to a well known and simple
geometry: the Schwarzschild solution. Initially we found non null
gravitomagnetic components in exact calculations, attributed to a
second order effect coming from the non-linearity of gravitation.
These components vanish when considering linearized gravity, making
in fact teleparallelism closer to electromagnetism. Concerning to
the gravitoelectric fields, they exhibited the expected form:
gravitoelectric components totally analogous to the Coulombian
electric field of electromagnetism. In this context, we have found
the gravitational field equations presenting strong similarity to
the Maxwell's electromagnetic ones.

A second test was performed with the purpose of making evident the
emergence of  gravitomagnetism  in linear regime, as in
electromagnetic case. The chosen geometry was, for a calculational
convenience and for easy comparison between theories, the massive
spherical shell in slow rotation. The result was surprising when
compared with the electromagnetic analogue (the charged spherical
shell in rotation), since they are quite similar.

There is another interesting point to be noted: the approach of
teleparallelism highlights the role played by the observers,
represented here by the tetrad fields. This subtleness is not
present in the metric description of gravity. Thus, even for a
static distribution of matter, it is in principle expected
gravitomagnetic field when considering observers in motion, that is
in complete agreement with electromagnetism. We can investigate, for
example, which kind of fields emerge in the context of a free
falling frame in the Schwarzschild spacetime, that is, a radial
accelerated frame (by the gravitational force) going straightly to
the singularity. Other interesting possibility is to consider an
observer with the same angular velocity of a rotating massive
spherical shell. These cases are under investigation.

Finally, we can say that the results obtained in the teleparallel
gravitomagnetism, although compatible to those of GR, present much
more affinity to electromagnetism from its first principles, based
on a gauge similar structure.

New tests for the model are desirable as well as other verifications
of theoretical consistency, since it opens real possibilities to
improve the comprehension of these two fundamental interactions and
has a strong relation with unification efforts.

\section{Acknowledgments}

\indent

The authors would like to thank J. W. Maluf for useful discussions
and R. F. P. Mendes for the  revision of the manuscript. They would
like also to thank CNPq for partial financial support.

\appendix

\section{Expanded first pair of gravitational field equation}\label{appA}
Equivalent to Gauss's law $(\rho=0)$:
\begin{eqnarray}\nonumber
& \partial _{i}(hE_{a}{}^{i}) - h[ {\cal H}^{b c}{}_{a i
j}E_{b}{}^{i}E_{c}{}^{j} + g_{ri}h^{c}{}_{j}\epsilon^{j r
k}(E_{c}{}^{i}B_{a k}-1/2E_{a}{}^{i}B_{c k}) \\ &+ {\cal I}^{b
c}{}_{a n i j}\epsilon^{j n k} E_{c}{}^{i}B_{b k} +{\cal J}^{c}{}_{i
j}E_{c}{}^{i}E_{a}{}^{j} +  {\cal K}^{b c}{}_{a r i j n}\epsilon^{i
j k}\epsilon^{n r t} B_{c k}B_{b t}]=0 \label{EQcampo1D2}
\end{eqnarray}
with
\begin{eqnarray}\nonumber
{\cal H}^{b c}{}_{a i j}= &-&g_{0
0}h_{a}{}^{0}h^{b}{}_{i}h^{c}{}_{j}+h_{a}{}^{0}g_{0
j}h^{b}{}_{i}{}h^{c}{}_{0} -\frac{1}{2}h_{a}{}^{0}g_{i
j}h^{b}{}_{0}h^{c}{}_{0} \\ &+& \nonumber h_{a}{}^{0}g_{0
i}h^{b}{}_{0}h^{c}{}_{j}+\frac{1}{2}h_{a}{}^{0}g_{0
0}h^{c}{}_{i}h^{b}{}_{j}-\frac{1}{2}h_{a}{}^{0}g_{0
j}h^{c}{}_{i}h^{b}{}_{0} \\&-&  \nonumber \frac{1}{2}h_{a}{}^{0}g_{0
i}h^{c}{}_{0}h^{b}{}_{j}+h_{a j}h^{c}{}_{i}h^{b}{}_{0}-2h_{a
j}h^{b}{}_{i}h^{c}{}_{0}+2h_{a 0}h^{b}{}_{i}h^{c}{}_{j} \\ &-&
\nonumber h_{a 0}h^{c}{}_{i}h^{b}{}_{j},
\\ \nonumber
{\cal I}^{b c}{}_{a n i j}&=&2h_{a}{}^{0}g_{0
n}h^{b}{}_{i}h^{c}{}_{j} - 2h_{a}{}^{0}g_{n
i}h^{b}{}_{0}h^{c}{}_{j}- h_{a}{}^{0}g_{0 n}h^{c}{}_{i}h^{b}{}_{j}
\\ &+& \nonumber h_{a}{}^{0}g_{n
i}h^{c}{}_{0}h^{b}{}_{j}+ h_{a n}h^{c}{}_{i}h^{b}{}_{j}-2h_{a
n}h^{b}{}_{i}h^{c}{}_{j},\nonumber
\\ \nonumber
{\cal J}^{c}{}_{i j}&=&g_{i j}h^{c}{}_{0}-2g_{0
 i}h^{c}{}_{j}+g_{0 j}h^{c}{}_{i},
\\ \nonumber
{\cal K}^{b c}{}_{a r i j n}&=&h_{a}{}^{0}g_{r
 i}h^{b}{}_{j}h^{c}{}_{n}+h_{a}{}^{0}\nonumber
g_{n i}h^{c}{}_{j}h^{b}{}_{r}.
\end{eqnarray}

Analogue to Amp\`ere's law $(\rho=q)$:
\begin{eqnarray}\nonumber &&
\epsilon^{q j k}\partial_{j}(hB_{a k})-\partial _{0}(hE_{a}{}^{q}) -
\,h[{\cal P}^{b c}{}_{a i}E_{b}{}^{i}E_{c}{}^{q}+ {\cal Q}^{b
c}{}_{a i j}\epsilon^{q i k} E_{b}{}^{j}B_{c k}
\\&+&\nonumber {\cal M}^{b c}{}_{i j a r}\epsilon^{r j
k}\epsilon^{q i t}B_{b k}B_{c t} + g_{r i}h^{c}{}_{j}\epsilon^{j r
k}\epsilon^{q i t}B_{a k}B_{c t} - \frac{1}{2}g_{r
i}h^{b}{}_{j}\epsilon^{j r k}\epsilon^{q i t}B_{b k}B_{a t}\\ &+&
\nonumber {\cal U}^{b c}{}_{i}h_{a j}E_{c}{}^{q}\epsilon^{i j k}B_{b
k}+{\cal V}^{b}{}_{i j}E_{b}{}^{j}\epsilon^{q i k}B_{a k}+ {\cal
W}^{c}{}_{i j}E_{a}{}^{j}\epsilon^{q i k}B_{c k} \\&-& \nonumber
g_{0 j}h^{c}{}_{i}\epsilon^{i j k}(E_{c}{}^{q}B_{a k}-
E_{a}{}^{q}B_{c k})+{\cal X}^{c}{}_{i}E_{a}{}^{i}E_{c}{}^{q}+
\frac{1}{2}{\cal Z}^{b}{}_{i}E_{b}{}^{i}E_{a}{}^{q}] \\ &+&
\nonumber hh_{a}{}^{q}[{\cal A}^{b c}{}_{i j}E_{c}{}^{i}E_{b}{}^{j}+
{\cal N}^{b c}{}_{n i j}\epsilon^{j n k}E_{c}{}^{i}B_{b k} + {\cal
C}^{b c}{}_{n i j}\epsilon^{i j k}E_{b}{}^{n}B_{c k} \\ &+& {\cal
D}^{b c}{}_{r i j n}\epsilon^{i j k}\epsilon^{n r t}B_{c k}B_{b
t}]=0 \label{EQcampoD22}
\end{eqnarray}
with
\begin{eqnarray}\nonumber
{\cal A}^{b c}{}_{i j}=&-&g_{0 0}h^{b}{}_{i}h^{c}{}_{j}+g_{0
j}h^{b}{}_{i}h^{c}{}_{0}- \frac{1}{2}g_{i
j}h^{b}{}_{0}h^{c}{}_{0}+g_{0 i}h^{b}{}_{0}h^{c}{}_{j} \\
&+& \nonumber \frac{1}{2}g_{0
0}h^{b}{}_{i}h^{c}{}_{j}-\frac{1}{2}g_{0 j}h^{c}{}_{i}h^{b}{}_{0}-
\frac{1}{2}g_{0 i}h^{c}{}_{0}h^{b}{}_{j},\nonumber
\\
{\cal P}^{b c}{}_{a i}&=&2h^{b}{}_{0}h^{c}{}_{0}h_{a
i}-2h^{b}{}_{0}h^{c}{}_{0}h_{a 0}+h^{c}{}_{0}h^{b}{}_{i}h_{a 0}-
h^{b}{}_{0}h^{c}{}_{0}h_{a i}, \nonumber
\\
{\cal Q}^{b c}{}_{a i j}&=&-2h^{b}{}_{i}h^{c}{}_{0}h_{a
j}+2h^{b}{}_{i}h^{c}{}_{j}h_{a 0}+h^{c}{}_{i}h^{b}{}_{0}h_{a j}-
h^{c}{}_{i}h^{b}{}_{j}h_{a 0},\nonumber
\\
{\cal N}^{b c}{}_{n i j}&=&g_{0 n}h^{b}{}_{i}h^{c}{}_{j}-g_{n
i}h^{b}{}_{0}h^{c}{}_{j}+ \frac{1}{2}g_{0
j}h^{c}{}_{i}h^{b}{}_{n}-\frac{1}{2}g_{i j}h^{c}{}_{0}h^{b}{}_{n},
\nonumber
\\
{\cal C}^{b c}{}_{n i j}&=&g_{n i}h^{b}{}_{j}h^{c}{}_{0}-g_{0
i}h^{b}{}_{j}h^{c}{}_{n}+ \frac{1}{2}g_{0
i}h^{c}{}_{j}h^{b}{}_{n}-\frac{1}{2}g_{n i}h^{c}{}_{j}h^{b}{}_{0},
\nonumber
\\
\nonumber {\cal D}^{b c}{}_{r i j n}&=&g_{r
i}h^{b}{}_{j}h^{c}{}_{n}+\frac{1}{2}g_{n
i}h^{c}{}_{j}h^{b}{}_{r},\nonumber
\\{\cal M}^{b c}{}_{i j a
r}&=&2h^{b}{}_{i}h^{c}{}_{j}h_{a r}-h^{b}{}_{j}h^{c}{}_{i}h_{a
r},\nonumber
\\
{\cal U}^{b
c}{}_{i}&=&2h^{b}{}_{0}h^{c}{}_{i}-h^{b}{}_{i}h^{c}{}_{0},\nonumber
\\
{\cal V}^{b}{}_{i j}&=&g_{0 i}h^{b}{}_{j}-g_{i
j}h^{b}{}_{0},\nonumber
\\{\cal W}^{c}{}_{i j}&=&2g_{i
j}h^{c}{}_{0}-2g_{0 i}h^{c}{}_{j},\nonumber
\\
{\cal X}^{c}{}_{i}&=&2g_{0 0}h^{c}{}_{i}-2g_{0
i}h^{c}{}_{0},\nonumber
\\
{\cal Z}^{b}{}_{i}&=&2g_{0 i}h^{b}{}_{0}-2g_{0
0}h^{b}{}_{i}.\nonumber
\end{eqnarray}
\section{Non vanishing torsion components for the spinning massive spherical shell}\label{appB}
\begin{eqnarray}\nonumber
T_{001}&=& T_{221} = T_{331} = - \frac{2\alpha x}{r^3},
\\ \nonumber
T_{002}&=& T_{112} =T_{332}= - \frac{2\alpha y}{r^3}, \\ \nonumber
T_{003}&=& T_{113} = T_{223} = - \frac{2\alpha z}{r^3},\\ \nonumber
T_{101}&=&-T_{202}= \left(\frac{2\alpha x}{r^3}\right) \left(
\frac{r_0^2(r_0 + 6\alpha)\Omega_0y}{r^3} \right),
\\ \nonumber
T_{103}&=&  \left(\frac{2\alpha z}{r^3}\right) \left(
\frac{r_0^2(r_0 + 6\alpha)\Omega_0y}{r^3} \right),
\\ \nonumber
T_{203} &=& - \left(\frac{2\alpha z}{r^3}\right) \left(
\frac{r_0^2(r_0 + 6\alpha)\Omega_0x}{r^3} \right),
\\ \nonumber
T_{201}&=& - \left(\frac{2\alpha x}{r^3}\right) \left(
\frac{r_0^2(r_0 + 6\alpha)\Omega_0x}{r^3} \right),
\\ \nonumber
T_{102}&=&  \left(\frac{2\alpha y}{r^3}\right) \left(
\frac{r_0^2(r_0 + 6\alpha)\Omega_0y}{r^3} \right),
\\ \nonumber
T_{013}&=& - \frac{3r_0^2\Omega_0yz}{r^5}\left( r_0 + 6\alpha -
\frac{2\alpha r_0}{r}\right) ,
\\ \nonumber
T_{023}&=& \frac{3r_0^2\Omega_0xz}{r^5}\left( r_0 + 6\alpha -
                           \frac{2\alpha r_0}{r}\right) ,
\\ \nonumber
T_{012}&=& \left( 1 - \frac{2\alpha}{r}\right)
\left(\frac{2r_0\Omega_0(r_0 + 6\alpha)}{r^3} -
\frac{3r_0^2\Omega_0(r_0 + 6\alpha)(x^2 + y^2)}{r^5}\right).
\end{eqnarray}

\end{document}